\def\be{\begin{equation}}
\def\ee{\end{equation}}
\def\ba{\begin{eqnarray}}
\def\ea{\end{eqnarray}}

\def\Npix{N_{\rm pix}}
\def\fsky{f_{\rm sky}}
\def\ellmax{\ell_{\rm max}}
\def\ta{\widetilde a}
\def\tC{\widetilde C}
\def\hC{\widehat C}
\def\tN{\widetilde N}
\def\eqdef{\stackrel{\rm def}{=}}
\def\bigoh{{\mathcal O}}
\def\simle{\lesssim}

\newcommand{\aut}[2]{{#2.\ #1,}}
\newcommand{\paut}[2]{{#2.\ #1} and}
\newcommand{\laut}[2]{{#2.\ #1,}}
\newcommand{\etal}{{\it et al. }}
\newcommand{\miscj}[4]{#1, {\bf #2}, #3 (#4).}
\newcommand{\ApJ}[3]{Astrophys.\ J., {\bf #1}, #2 (#3).}

\newcommand{\MNRAS}[3]{Mon.\ Not.\ R.\ Astron.\ Soc., {\bf #1}, #2 (#3).}
\newcommand{\PRD}[3]{Phys.\ Rev.\ D, {\bf #1}, #2 (#3).}
\newcommand{\PRL}[3]{Phys.\ Rev.\ Lett., {\bf #1}, #2 (#3).}
\newcommand{\pp}[1]{preprint, astro-ph/#1.}

\newcommand{\astroph}[1]{}

\documentclass{elsart}
\journal{New Astronomy}
\usepackage{natbib}

\usepackage{graphicx}
\usepackage{epsfig}

\usepackage{amssymb}

\begin{document}

\begin{frontmatter}



\title{Pure pseudo-$C_\ell$ estimators for CMB B-modes}


\author{Kendrick M. Smith}
\address{Kavli Institute for Cosmological Physics, Enrico Fermi Institute, University of Chicago, 60637\\
Department of Physics, University of Chicago, 60637}

\begin{abstract}
Fast heuristically weighted, or pseudo-$C_\ell$, estimators are a
frequently used method for estimating power spectra in CMB surveys with large numbers of pixels.
Recently, Challinor \& Chon showed that the E-B mixing 
in these estimators can become a dominant contaminant at low
noise levels, ultimately limiting the gravity wave signal which can
be detected on a finite patch of sky.
We define a modified version of the estimators which eliminates
E-B mixing and is near-optimal at all noise levels.
\end{abstract}


\end{frontmatter}

% main text
\section{Introduction}

Measurements of CMB polarization, reported by four experiments to date
\cite{DASI1,CBI1,CAPMAP,Boom03}, are rapidly improving in sensitivity.
Future ground-based surveys will be optimized for sensitivity to the
B-mode power spectrum, which offers a unique window into the physics of 
the early universe via the gravity wave signal \cite{SZgw,KKSgw},
and can help break parameter degeneracies via the lensing signal \cite{Stompor:1998zj,me2}.

On a practical level, two general methods for estimating power spectra from noisy
maps have been used in experiments to date: maximum likelihood \cite{BJK}, and
pseudo-$C_\ell$ estimators \cite{WHGCl,MASTER,HGpolCl}.
The strengths and weaknesses of these methods are complementary.
Maximum likelihood power spectrum estimatation is optimal (in the sense that the Cramer-Rao
inequality is saturated) but computationally expensive, requiring $\bigoh(\Npix^3)$
time and $\bigoh(\Npix^2)$ memory, and rapidly becomes infeasible for large maps.
In contrast, pseudo-$C_\ell$ estimators are very fast, suboptimal in principle, but frequently
near-optimal in practice.  For this reason, they are currently the method of choice for surveys with
large pixel counts.

In the context of CMB polarization, pseudo-$C_\ell$ estimators have an additional drawback,
as studied by \citet{ChalChon}: the estimated BB power spectrum acquires a nonzero 
contribution from E-modes in the map.
The pseudo-$C_\ell$ construction includes a debiasing step (Eq.~(\ref{eq:debias}) below) which removes
the E-B mixing in the mean; however, the variance of the BB estimators still depends on the
level of EE power.
This is analagous to treating noise by subtracting the noise bias from each estimator; even 
though the bias from noise is removed in the mean, it still makes a contribution to the
estimator variance.
This contribution can dominate if the instrumental noise is sufficiently small; e.g. in \cite{ChalChon}
it is shown that for surveys with $\fsky \sim 0.01$, it limits the gravity wave signal which can be
detected using pseudo-$C_\ell$ estimators to $T/S\sim 0.05$.

We will describe a modification to the pseudo-$C_\ell$ construction, defining {\em pure pseudo-$C_\ell$ estimators} which 
separate E and B in the strongest possible sense.
On a finite patch of sky, the BB estimator completely filters out E-modes, acquiring
contributions only from the B-mode signal and noise.
We will show that these estimators significantly improve power spectrum errors for noise levels
$\simle 20$ $\mu$K-arcmin, and impose no limit on the gravity wave signal which can be detected.

\section{The pseudo-$C_\ell$ construction}
\label{sec:pcl}

In this section, we briefly review the pseudo-$C_\ell$ construction before presenting our modification in \S\ref{sec:eb}.
For more details, see \cite{HGpolCl}.

The basic idea is easy to explain.  One constructs power spectrum estimators on a finite patch of sky
by weighting the observed polarization map $\Pi_{ab}(x)$ by a heuristically chosen weight function $W(x)$,
then computing $a_{\ell m}$'s and power spectra as if the weighted polarization were an all-sky field:
\ba
\ta^E_{\ell m} &\eqdef& \int d^2x\, 2 W(x) \Pi^{ab}(x) Y^{E*}_{(\ell m)ab}(x) \\
\ta^B_{\ell m} &\eqdef& \int d^2x\, 2 W(x) \Pi^{ab}(x) Y^{B*}_{(\ell m)ab}(x) \label{eq:defblm} \\
\tC^{EE}_\ell &\eqdef& \frac{1}{2\ell+1} \sum_m a^{E*}_{\ell m} a^E_{\ell m}  \qquad ; \qquad
   \tC^{BB}_\ell \eqdef \frac{1}{2\ell+1} \sum_m a^{B*}_{\ell m} a^B_{\ell m}  \label{eq:defclbb}
\ea
where $Y^E_{\ell m}$, $Y^B_{\ell m}$ denote E-mode and B-mode spherical harmonics \cite{KKS,ZSspins}.
We have introduced the notation $\tC_\ell$ for the power spectrum of the {\em weighted}
polarization field, or ``pseudo power spectrum''.
Because multiplication by the weight function in real space mixes multipoles in harmonic space, the
pseudo power spectrum $\tC_\ell$ acquires contributions from multipoles $\ell' \ne \ell$.  In the mean,
this contribution can be written in terms of transfer matrices $K_{\ell\ell'}^\pm$:
\be
\left( \begin{array}{c} \langle \tC_\ell^{EE} \rangle \\ \langle \tC_\ell^{BB} \rangle \end{array}  \right) =
\left( \begin{array}{cc} K_{\ell\ell'}^+ &  K_{\ell\ell'}^- \\
                         K_{\ell\ell'}^- &  K_{\ell\ell'}^+ \end{array} \right)
\left( \begin{array}{c} C_{\ell'}^{EE} \\ C_{\ell'}^{BB} \end{array}  \right) + 
\left( \begin{array}{c} \tN_\ell^{EE} \\ \tN_\ell^{BB} \end{array}  \right)
\ee
where the vectors $\tN_\ell$ represent noise bias.
Note that the matrices $K^-_{\ell\ell'}$, $K^+_{\ell\ell'}$ represent multipole mixing
from the weight function with and without accompanying mixing of E and B.

The final ingredient in the pseudo-$C_\ell$ construction is a debiasing step:
\be
\left( \begin{array}{c} \hC_\ell^{EE} \\ \hC_\ell^{BB} \end{array}  \right) \eqdef
\left( \begin{array}{cc} K_{\ell\ell'}^+ &  K_{\ell\ell'}^- \\
                         K_{\ell\ell'}^- &  K_{\ell\ell'}^+ \end{array} \right)^{-1}
\left( \begin{array}{c} \tC_{\ell'}^{EE} - \tN_{\ell'}^{EE}  \\  \tC_{\ell'}^{BB} - \tN_{\ell'} \end{array}  \right) \label{eq:debias}
\ee
This removes the bias imposed by the weight function and noise; each $\hC_\ell$
is an unbiased power spectrum estimator.
(As a technical point, we mention that the matrix inversion in Eq.~(\ref{eq:debias}) can only be
performed if $\fsky$ is large; otherwise one has to bin multipoles into bandpowers.)

\section{E-B mixing}
\label{sec:eb}

An elegant framework for studying E-B mixing has been presented in \cite{BunnPure}, which 
was the original inspiration for this work.
On a finite patch of sky, the polarization field can be decomposed into three types of modes:
{\em pure E-modes}, which receive no contribution from the B-mode signal,
{\em pure B-modes}, which likewise receive no contribution from the E-mode signal,
and {\em ambiguous modes}, which receive contributions from both.

In terms of this decomposition,
the underlying reason why pseudo-$C_\ell$ estimators mix E and B modes on a finite patch of
sky can be understood as follows.
Returning to the definition in Eq.~(\ref{eq:defblm}), the pseudo multipole $\ta^B_{\ell m}$ is given by the
overlap integral with a mode (in brackets below) which is a mixture of all three types:
\be
\ta^B_{\ell m} = \int d^2x\, 2 \Pi^{ab}(x) \Big[ W(x) Y^{B*}_{\ell m}(x) \Big]  \label{eq:eb1}
\ee
Therefore, the E-mode signal makes a nonzero contribution to $\ta^B_{\ell m}$, via the pure E-mode and ambiguous mode
components of the mode in brackets.
In this way, the E-mode signal finds its way into the BB power spectrum estimator $\hC^{BB}_\ell$,
where it contributes extra variance.

Ideally, one would like to have a BB power spectrum estimator which does not receive
contaminating contributions from the larger E-mode signal.
This suggests modifying the definition in Eq.~(\ref{eq:eb1}) so that the mode in brackets is a pure B-mode.

According to \cite{BunnPure}, pure B-modes on a finite patch of sky are of the form
\be
\left( \frac{1}{2} \epsilon_{ac}\nabla^c\nabla_b + \frac{1}{2} \epsilon_{bc}\nabla^c\nabla_a \right) \Phi(x)  \label{eq:eb2}
\ee
where $\Phi(x)$ is a potential which must satisfy the boundary conditions $\Phi = \nabla_a\Phi = 0$.
To make contact with the pseudo multipole in Eq.~(\ref{eq:eb1}), we write the B-mode spherical harmonic $Y^B_{\ell m}$ as
the result of applying the same differential operator to the scalar spherical harmonic $Y_{\ell m}$:
\be
\ta^{B}_{\ell m} = N_\ell \int d^2x\, 2 \Pi^{ab}(x) \left[
 W(x) \left( \frac{1}{2} \epsilon_{ac}\nabla^c\nabla_b + \frac{1}{2} \epsilon_{bc}\nabla^c\nabla_a \right) Y^*_{\ell m}(x)
\right]   \label{eq:eb3}
\ee
where $N_\ell=1/\sqrt{(\ell-1)\ell(\ell+1)(\ell+2)}$.
Comparing Eqs.~(\ref{eq:eb2}) and~(\ref{eq:eb3}), it is natural to modify the definition of $\ta^B_{\ell m}$
by bringing the weight function inside the differential operator.
We take this as the definition of the {\em pure pseudo multipole:}
\be
\ta^{B,pure}_{\ell m} \eqdef N_\ell \int d^2x\, 2 \Pi^{ab}(x) \left[
  \left( \frac{1}{2} \epsilon_{ac}\nabla^c\nabla_b + \frac{1}{2} \epsilon_{bc}\nabla^c\nabla_a \right) W(x) Y^*_{\ell m}(x)
\right]  \label{eq:defpblm}
\ee
For the mode in brackets to be a pure B-mode, the weight function must satisfy
the boundary conditions $W = \nabla_a W = 0$.  We will return to this in \S\ref{sec:statweight}.

We define {\em pure pseudo-$C_\ell$ estimators} by replacing $\ta^B_{\ell m} \rightarrow \ta^{B,pure}_{\ell m}$
and leaving subsequent steps in the pseudo-$C_\ell$ construction from \S\ref{sec:pcl} unchanged.
When this modification is made, the $E\rightarrow B$ term in the transfer matrix is always zero:
\be
\left( \begin{array}{cc} K_{\ell\ell'}^+ &  K_{\ell\ell'}^- \\
                         K_{\ell\ell'}^- &  K_{\ell\ell'}^+ \end{array} \right)
\rightarrow
\left( \begin{array}{cc} K_{\ell\ell'}^+ &  K_{\ell\ell'}^- \\
                                0        &  K_{\ell\ell'}^{+,pure} \end{array} \right)
\ee
This corresponds to the statement that pure pseudo-$C_\ell$ estimators do not mix $E\rightarrow B$ in
the mean.  The estimators also separate E and B in a much stronger sense: in a single realization, the
estimated BB power receives zero contribution from E-modes in the map.
This is because the observed polarization $\Pi^{ab}$ always overlaps a pure B-mode (in brackets in Eq.~(\ref{eq:defpblm}))
which by construction completely filters out E-modes, even on a finite patch of sky.
Because of this strong separation, the pure pseudo-$C_\ell$ estimator is completely ``blind'' to E-modes
and EE power does not act as an source of extra variance.

An essential ingredient in the pseudo-$C_\ell$ construction is fast ($\bigoh(\ellmax^3)$)
algorithms for evaluating the estimators and precomputing the transfer matrix.
Variants of these algorithms can also be given for pure pseudo-$C_\ell$ estimators \cite{me},
but we omit the details here.

\section{Statistical weight}
\label{sec:statweight}

An important practical issue, for both pure and ordinary pseudo-$C_\ell$ estimators, is choosing
the pixel weight function $W(x)$.
In general, this must be done empirically (perhaps guided by heuristics) using Monte Carlo simulations
to optimize the estimator variance.
The optimal weight function will depend on $\ell$; at low $\ell$ or high signal-to-noise, uniform
weighting (which minimizes sample variance) is near-optimal, whereas at high $\ell$ or low
signal-to-noise, inverse noise weighting (which minimizes noise variance) is near-optimal \cite{ChalChon}.

For pure pseudo-$C_\ell$ estimators, an extra complication arises when choosing the weight function.
The statistical weight of a pixel is given by a combination of the weight function $W(x)$ and its
first two derivatives in the pixel.
This can be seen from Eq.~(\ref{eq:defpblm}), where the estimator is defined by overlapping the map $\Pi^{ab}(x)$ with
a pure B-mode (in brackets) which contains terms proportional to $W$, its first derivative, and second derivative.
In contrast, for ordinary pseudo-$C_\ell$ estimators, the map overlaps a mode which is proportional to $W(x)$,
and the statistical weight of a pixel is simply given by the weight function.

\begin{figure}
\centerline{ \epsfxsize=3.0truein\epsffile{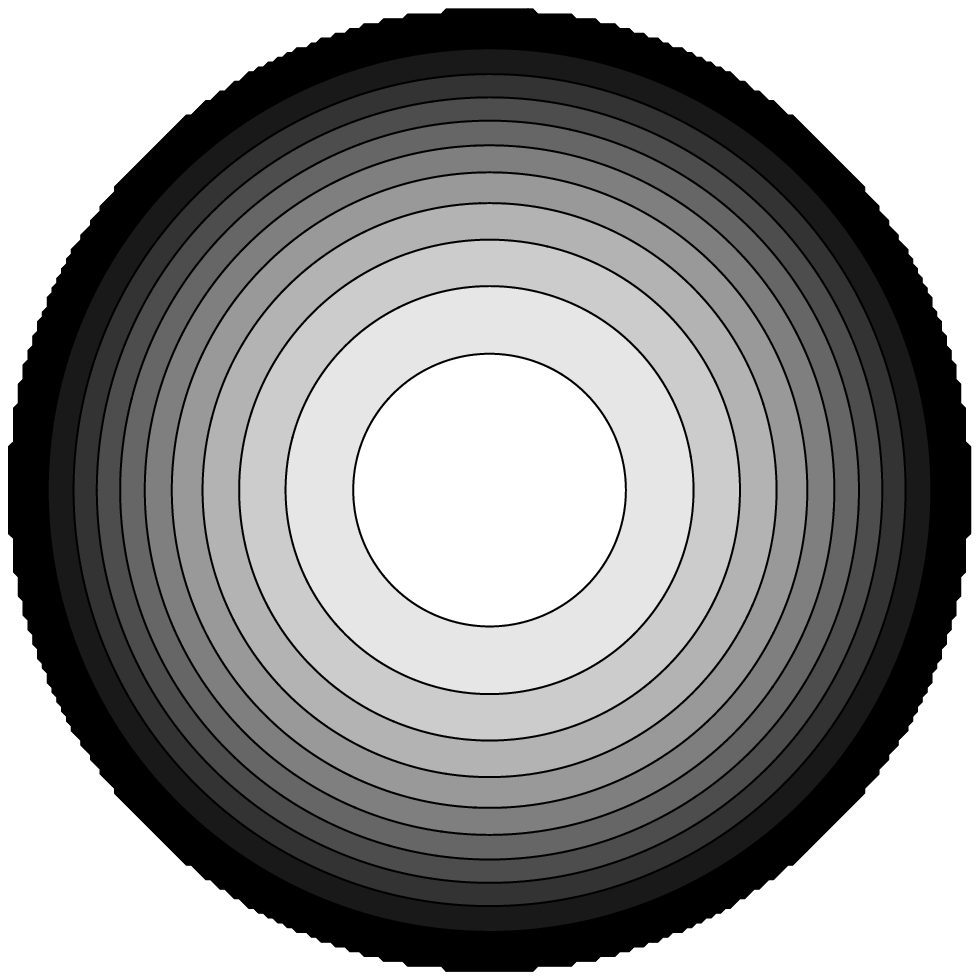} \epsfxsize=3.0truein\epsffile{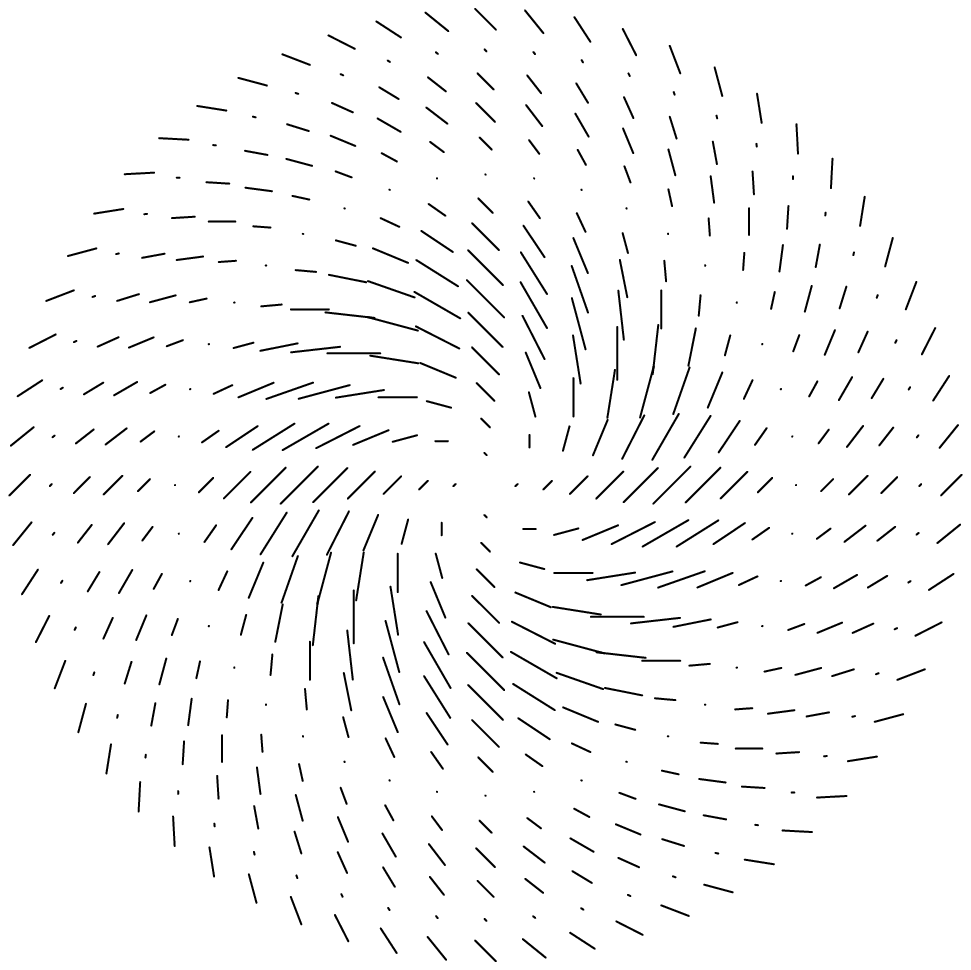} }
\caption{Weight function $W(x)$, satisfying the boundary conditions $W = \nabla_a W = 0$ required for
pure pseudo-$C_\ell$ power spectrum estimation (left panel) and the resulting pure B-mode defined by Eq.~(\ref{eq:pureb})
(right panel).}
\label{fig:bessel}
\end{figure}

As a concrete example (taken from Ref.~\cite{ZS}), the weight function $W(x)$ shown in Fig.~\ref{fig:bessel}, left panel, is 
optimized for pure pseudo-$C_\ell$ power spectrum estimation at $\ell=30$ on a circular patch of sky with radius $13^\circ$
and uniform white noise.
If this weight function were used in ordinary pseudo-$C_\ell$ estimation, the statistical weight
would be concentrated near the center of the survey and the resulting estimator would be suboptimal
(since the noise distribution is assumed uniform).
For pure pseudo-$C_\ell$ estimators, one can understand the distribution of statistical weight by
plotting the pure B-mode
\be
\left( \frac{1}{2} \epsilon_{ac}\nabla^c\nabla_b + \frac{1}{2} \epsilon_{bc}\nabla^c\nabla_a \right) W(x) Y^*_{\ell m}(x) \label{eq:pureb}
\ee
which overlaps the map $\Pi^{ab}(x)$ in Eq.~(\ref{eq:defpblm}).
This is shown, for $(\ell,m)=(30,0)$, in the right panel of Fig.~\ref{fig:bessel}.
It is seen that the statistical weight is distributed roughly evenly throughout the survey.
(For the $m=0$ mode shown here, the statistical weight is somewhat biased toward the center of the region,
but would be uniformly distributed after averaging over $m$ as in Eq.~(\ref{eq:defclbb}).)

As remarked in \S\ref{sec:eb}, a general feature of pure pseudo-$C_\ell$ estimation is that the weight function 
must be apodized so that the boundary conditions $W = \nabla_a W = 0$ are satisfied.
Suppose that one tried to ``cheat'' by using a weight function which is constant
throughout the survey except for a tiny ribbon around the boundary, where it rapidly
goes to zero in order to satisfy the boundary conditions.
Then the resulting estimator would have most of its statistical weight in the ribbon, where the
derivatives become large, and would therefore be very suboptimal.
In fact, one can think of the boundary conditions $W = \nabla_a W = 0$ as a consequence of
the statistical weight being given by the first two derivatives.  If the boundary conditions
were not satisfied, then the estimator still makes sense mathematically but the statistical
weight would include delta functions on the boundary, and does not make sense for noisy data.

Let us emphasize that, although pure pseudo-$C_\ell$ estimators require the weight function
$W(x)$ to be apodized near the boundary, the distribution of statistical weight need not be
apodized.
Any distribution of statistical weight can be obtained by choosing $W(x)$ such that the boundary
conditions are satisfied, and the amplitude of the pure B-mode in Eq.~(\ref{eq:pureb}) matches the desired distribution, 
For example, the weight function considered in this section has been engineered to give
uniform statistical weight (to match the assumption of uniform noise).
This example is part of a systematic framework for optimizing weight functions for both
pure and ordinary pseudo-$C_\ell$, which will be presented separately \cite{ZS}.

\section{Examples}

\begin{figure}
\centerline{\epsfxsize=6.0truein\epsffile[18 450 592 710]{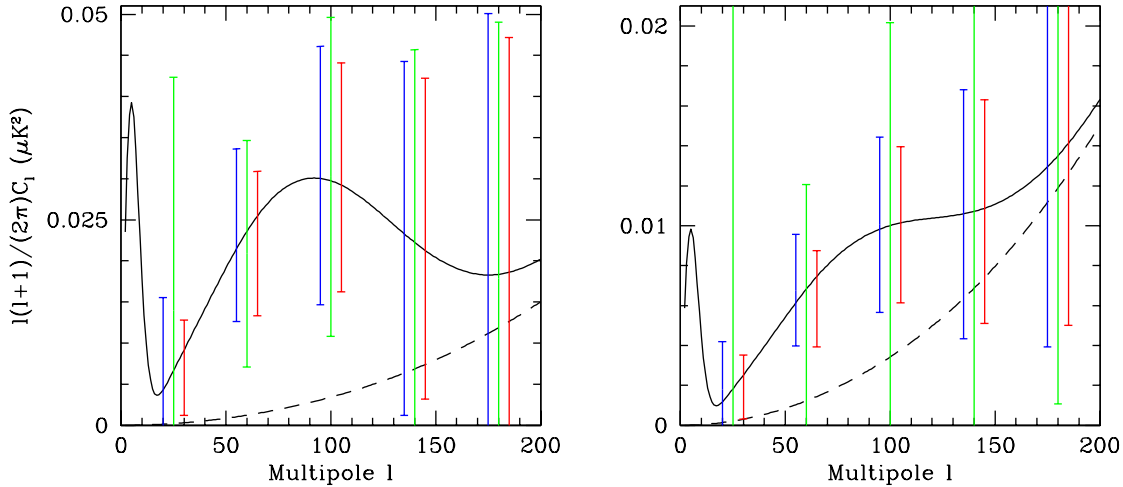}}
\caption{ BB power spectrum errors for pure pseudo-$C_\ell$ estimators (blue/left),
ordinary pseudo-$C_\ell$ estimators (green/middle), and the Cramer-Rao bound (red/right).
In the left panel, we have shown a model with $T/S=0.2$ and noise level 20 $\mu$K-arcmin;
in the right panel, a model with $T/S=0.05$ and noise 10 $\mu$K-arcmin.  The dashed lines
show the lensing component of the BB power spectrum; detecting the gravity wave signal
requires measuring power in excess of this level. }
\label{fig:example}
\end{figure}

We now consider some results for a mock survey
on a spherical cap with radius $13^\circ$, with uniform
white noise and a Gaussian beam with $\theta_{\rm FWHM}=25$ arcmin.
In Fig.~\ref{fig:example}, we compare the power spectrum errors obtained using
pure pseudo-$C_\ell$ estimators, unmodified pseudo-$C_\ell$ estimators, and
the Cramer-Rao bound on the estimator variance.

It is seen that the pure pseudo-$C_\ell$ construction significantly improves
power spectrum errors, relative to unmodified pseudo-$C_\ell$, for noise levels 
$\simle 20$ $\mu$K-arcmin.
For these low noise levels, the extra variance in unmodified pseudo-$C_\ell$
estimators can limit the gravity wave signal which can be detected.
For example, in the right panel (10 $\mu$K-arcmin), a gravity wave signal with $T/S=0.05$ 
can be detected using pure pseudo-$C_\ell$ estimators, but not using the unmodified versions.

\begin{figure}
\centerline{\epsffile[20 165 400 425]{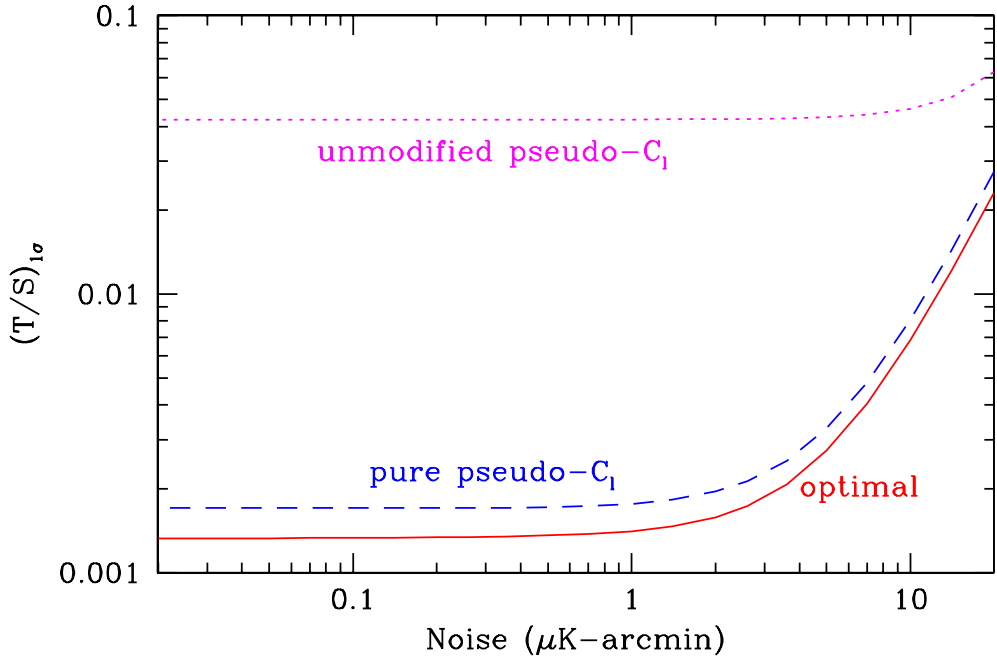}}
\caption{ Minimum $T/S$ detectable at $1\sigma$, for a spherical cap shaped survey with radius $13^\circ$,
uniform white noise, and 25 arcmin beam.}
\label{fig:ts}
\end{figure}

In Fig.~\ref{fig:ts}, we have shown the values of $T/S$ which can be detected at $1\sigma$ using each
estimator, for a variety of noise levels.
We find that the smallest gravity wave signal which can be detected using unmodified pseudo-$C_\ell$
estimators is $T/S=0.042$, in agreement with \citet{ChalChon}, but that the pure versions are $\sim 80$\%
optimal all the way to the floor at $T/S\sim 10^{-3}$, which represents the limit from the lensing component 
of the BB power spectrum (treated as a Gaussian contaminant) for this survey region.

\section{Discussion}

For CMB surveys with many pixels, pseudo-$C_\ell$ power spectrum estimation is currently the ``industry standard'',
but has a shortcoming for future high-sensitivity polarization experiments: the EE signal acts as an extra
source of noise on the BB power spectrum, becoming a significant contaminant at noise levels $\simle 20$ $\mu$K-arcmin.
We have defined pure pseudo-$C_\ell$ estimators, which eliminate this shortcoming while preserving the fast 
($\bigoh(\ellmax^3)$) evaluation of the estimator and transfer matrix.
These estimators completely filter out the E-mode component of the map; the estimated BB power receives contributions
only from the B-mode part of the signal and from noise.

A property of the pure pseudo-$C_\ell$ estimator is that the statistical weight is given
by a combination of the weight function $W(x)$ and its first two derivatives.
This imposes no restriction on the way the statistical weight is distributed (one can always choose
$W(x)$ to give any desired statistical weight while satisfying the boundary conditions) but makes 
optimizing the weight function for a given noise distribution less intuitive.
The optimization problem will be studied systematically in \cite{ZS}, in the context
of both pure and ordinary pseudo-$C_\ell$ estimators.

We thank Niayesh Afshordi, Wayne Hu, Dragan Huterer, Bruce Winstein, and Matias Zaldarriaga
for collaborative work and stimulating discussions.
This work was supported by the Kavli Institure for Cosmological Physics through the grant NSF PHY-0114422.

\end{document}